\documentclass[fleqn,usenatbib]{mnras}

\usepackage{newtxtext,newtxmath}

\usepackage[T1]{fontenc}

\DeclareRobustCommand{\VAN}[3]{#2}
\let\VANthebibliography\thebibliography
\def\thebibliography{\DeclareRobustCommand{\VAN}[3]{##3}\VANthebibliography}

\usepackage{ae,aecompl}
\usepackage{graphicx}	
\usepackage{amsmath}	
\usepackage{times}
\usepackage[usenames,dvipsnames]{xcolor}
\usepackage{soul}
\usepackage{comment}
\usepackage{tablefootnote}
\usepackage{threeparttable}

\title[Substellar IMF in NGC 2024]{JWST's Constraints on the Substellar IMF in NGC 2024}

\author[K. L. Luhman]{
K. L. Luhman$^{1,2}$\thanks{E-mail: kll207@psu.edu}
\\
$^{1}$Department of Astronomy and Astrophysics, The Pennsylvania State University, University Park, PA 16802, USA\\
$^{2}$Center for Exoplanets and Habitable Worlds, The Pennsylvania State University, University Park, PA 16802, USA
}

\date{Accepted 2026 July 17. Received 2026 May 24; in original form 2026 April 12}

\pubyear{2026}

\begin{document}
\label{firstpage}
\pagerange{\pageref{firstpage}--\pageref{lastpage}}
\maketitle

\begin{abstract}
A recent study has reported the detection of a turnover in the initial
mass function (IMF) below 12~$M_{\rm Jup}$ and the possible detection
of a minimum mass near 3~$M_{\rm Jup}$ in the heavily embedded cluster 
NGC 2024 ($<1$~Myr), which was based on a sample of brown 
dwarf candidates in NIRCam images from the James Webb Space Telescope (JWST).
I have used those NIRCam data in conjunction with archival
spectra from JWST's Near-Infrared Spectrograph (NIRSpec) to constrain the
substellar IMF in NGC 2024. I present NIRSpec data for 87 sources, 67
of which are within the NIRCam field. Based on those spectra and 
data from previous studies (e.g., X-ray emission), I have classified 
45 of the NIRSpec targets as members of NGC 2024, 17 of which have
spectral types indicative of brown dwarfs (M6.5--L; 13 are within NIRCam). 
The latter have mass estimates as low as $\sim4$~$M_{\rm Jup}$ according to
theoretical evolutionary models. 
There remain a few photometric candidates lacking spectra that extend down 
to the completeness limit of the images ($\gtrsim$2~$M_{\rm Jup}$ for 
$A_K\gtrsim1$), so the minimum mass of the IMF has not been detected. 
It is not possible to measure a reliable substellar IMF from the NIRCam images
because of the small number of brown dwarfs encompassed by them,
a bias against objects at lower masses due to the high extinctions,
and incompleteness at $\gtrsim10$~$M_{\rm Jup}$ due to the saturation limit.
Thus, there is no evidence in the JWST data for a turnover below 
12~$M_{\rm Jup}$.
\end{abstract}




\section{Introduction}
\label{sec:intro}

For three decades, nearby star-forming regions have served
as valuable targets for surveys to identify brown dwarfs at progressively
lower masses \citep{hil97,luh97,bri98,bej99,zap99}. One of the
primary objectives of that work has been the measurement of the shape and
minimum mass of the initial mass function (IMF) of brown dwarfs.
The challenge for brown dwarf surveys in star-forming regions lies in both 
the detection of these faint objects in imaging and distinguishing 
them from contaminants in the form of field stars and galaxies. 
The identification of brown dwarf candidates is more difficult 
in young populations with higher levels of extinction from their natal
molecular clouds, which reduces the available bands of photometry and increases 
the overlap in colors between brown dwarfs and reddened background sources. 
Even when relatively low extinction ($A_V<5$) allows good discrimination 
between brown dwarfs and most field stars and galaxies, samples of brown dwarf 
candidates remain subject to contamination from field brown dwarfs and 
galaxies with unusual colors \citep{luh25} (hereafter L25).  As a result, to 
derive a reliable substellar IMF for a star-forming region, spectroscopy or 
proper motion measurements (preferably both) are essential for confirming 
membership of brown dwarf candidates.

As the most sensitive infrared (IR) telescope to date, the James Webb 
Space Telescope \citep[JWST,][]{gar23} has been used to search for
brown dwarfs in a few nearby star-forming clusters that are sufficiently 
compact for its small field of view \citep[][L25]{mcc23,luh24ic,lan24}.
One of the targets for JWST has been NGC 2024, which is is a heavily 
obscured cluster in the Orion-B molecular cloud 
\citep{com96,mey96,mey08,rob24,rot26}. It has an age of $<1$~Myr 
\citep{lev06,get14b} and a distance of 403$\pm$5~pc \citep{dzi26}.
\citet{def25} (hereafter D25) obtained deep images in the center of 
NGC 2024 using NIRCam on JWST \citep{rie05,rie23}. They reported the detection
of brown dwarf candidates that indicated a turnover in the IMF 
below 12~$M_{\rm Jup}$ and a possible minimum mass near 3~$M_{\rm Jup}$,
However, none of the candidates were observed with spectroscopy to
confirm their nature as young brown dwarfs.  In addition, D25 did not provide 
a tabulation of their brown dwarf candidates, so it is not possible for 
subsequent studies to test their validity or reproduce the reported IMF. 
L25 presented a brief analysis of the images from D25, using a color-color 
diagram to identify eight candidates for brown dwarfs with spectral types 
of $\geq$L0.

JWST program 5409 (PI: M. De Furio) has used the multiobject mode of the
Near-Infrared Spectrograph \citep[NIRSpec,][]{fer22,jak22} to obtain
low-resolution 1--5~\micron\ spectra of sources in the center of NGC 2024,
many of which are located within the NIRCam images from D25.
In this paper, I use those NIRSpec data and the photometry from NIRCam
to constrain the substellar IMF in NGC 2024.

\section{JWST/NIRSpec Observations}

JWST program 5409 observed $\sim$100 sources toward the center of NGC 2024
using the multiobject spectroscopy mode of NIRSpec on 2025 February 10 (UT).
That mode employs the microshutter assembly (MSA), which consists of
four quadrants that each contain 365$\times$171 shutters.
A given shutter has a size of $0\farcs2\times0\farcs46$.
The MSA covers an area of $3\farcm6\times3\farcm4$.
NIRSpec utilizes two $2048\times2048$ detector arrays that have pixel 
sizes of $0\farcs103\times0\farcs105$. The observations in NGC 2024 were 
performed with four MSA configurations and the PRISM disperser, which covers
0.6--5.3~\micron\ with a spectral resolution of $\sim$40 to 300 from 
shorter to longer wavelengths. In each MSA configuration, data were 
taken at three nod positions in three adjacent shutters.
Each group of three shutters was equivalent to a $0\farcs2\times1\farcs5$ 
slitlet. The exposure at each nod position utilized 23 groups, 
five integrations, and the NRSIRS2RAPID readout pattern.
The total exposure time for each MSA configuration was 5252~s.

I retrieved the {\tt uncal} files for program 5409 from the Mikulski Archive 
for Space Telescopes (MAST)\footnote{\url{https://doi.org/10.17909/fyyq-st11}}.
Those files were processed with the JWST Science Calibration pipeline
version 1.20.2\footnote{\url{https://doi.org/10.5281/zenodo.17515973}}.
For each of the three nods for a given target, I performed three versions
of the background subtraction that utilized both of the other two nods
and each of those nods separately. 
I selected the option that minimized the residuals from
the nebular emission lines. I then compared the best background-subtracted
spectra among the three nods, and omitted the spectra for some nods for which 
the subtraction was too poor because of the spatial variations of the
nebular lines or neighboring sources appearing within the slitlits.
The spectra with adequate background subtraction among the three nods 
were combined. Some of the final spectra exhibit positive or negative
residuals at the wavelengths of nebular emission lines due to imperfect
background subtraction (e.g., hydrogen lines and the PAH feature
at 3.3~\micron), so those features should be treated with caution.
The data for several targets were not useful because of saturation.
I present reduced spectra for 87 sources ($m_{444}\sim8$--19), 
which are available in the online supplemental material. 
Some of the spectra have incomplete wavelength 
coverage because of truncation by the edges of the detectors.

\begin{figure*}
\includegraphics[width=0.95\textwidth]{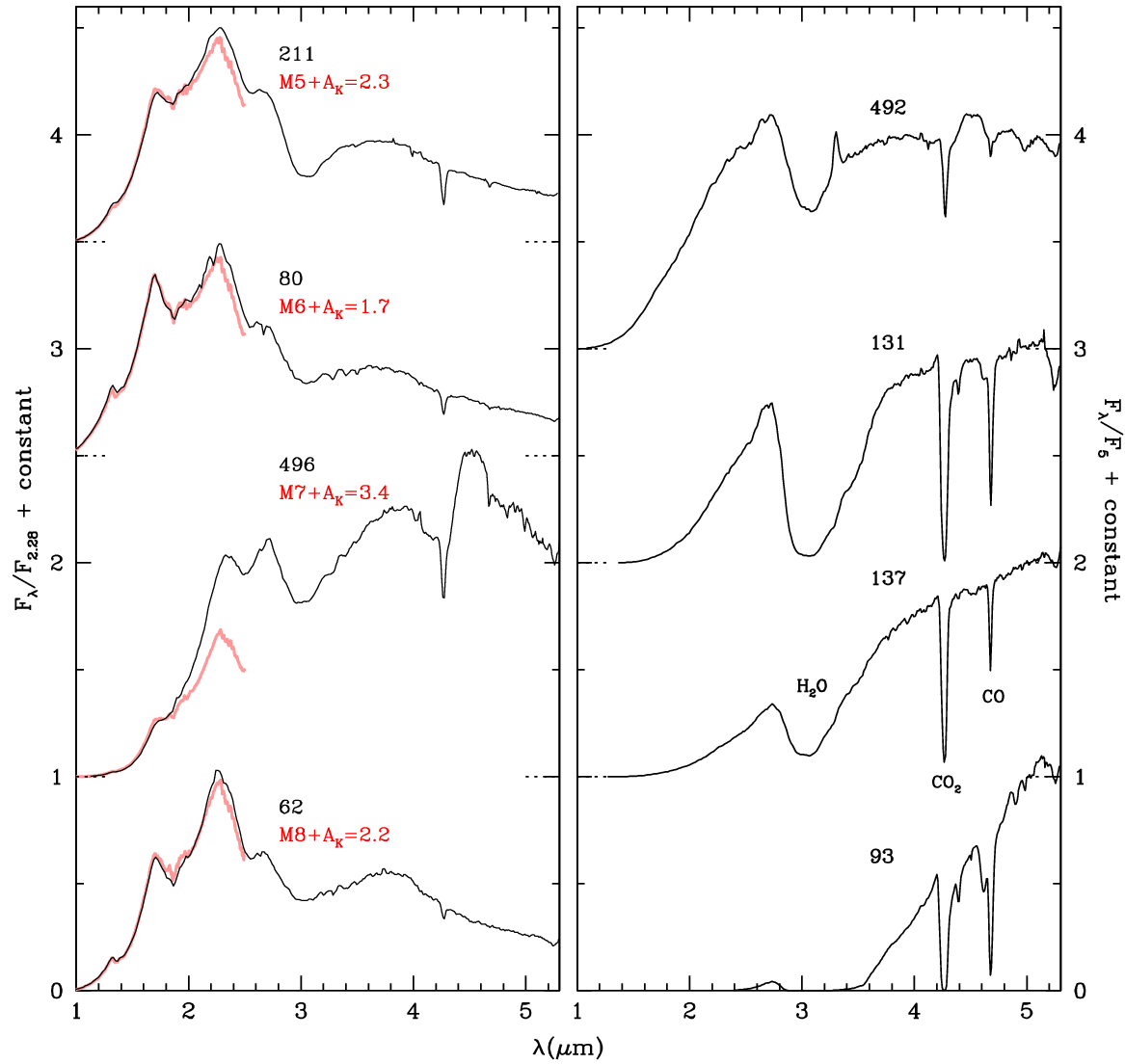}
\caption{Examples of JWST/NIRSpec spectra of sources classified as members of
NGC 2024 (black). They are labeled with the source numbers from the APT catalog 
for these observations (Table~\ref{tab:spec}). The spectra in the left panel 
are compared to young standard spectra that have been reddened to match the 
slopes from 1--1.7~\micron\ \citep[red,][]{luh17}. The absorption features at 
$\gtrsim3$~\micron\ are from interstellar ices (H$_2$O, CO$_2$, CO), as
indicated for source 137. For each spectrum that has been shifted upward, 
the level for zero flux is marked by a pair of dotted tick marks.}
\label{fig:spec1}
\end{figure*}

\begin{table}
\centering
\caption{Spectral and membership classifications and NIRCam photometry
for NIRSpec targets in NGC 2024. The format and content are described here. 
The full table is available in the online supplemental material.}
\label{tab:spec}
\begin{threeparttable}
\begin{tabular}{ll}
\hline
Column Label & Description\\
\hline
ID & Source number in catalog for JWST program 5409\\
RAdeg & Right ascension (ICRS)\\
DEdeg & Declination (ICRS)\\
Ref-Pos & Reference for right ascension and declination\tnote{\textit{a}} \\
SpType & Spectral type from NIRSpec\tnote{\textit{b}}\\
ak & $A_K$ from NIRSpec\\
E\_ak & Upper error in ak\\
e\_ak & Lower error in ak\\
member & NGC 2024 member?\\
evidence & Evidence of Membership\\
m115mag & F115W NIRCam magnitude\tnote{\textit{c}} \\
e\_m115mag & Error in m115mag \\
m140mag & F140M NIRCam magnitude\tnote{\textit{c}} \\
e\_m140mag & Error in m140mag \\
m182mag & F182M NIRCam magnitude\tnote{\textit{c}} \\
e\_m182mag & Error in m182mag \\
m360mag & F360M NIRCam magnitude\tnote{\textit{c}} \\
e\_m360mag & Error in m360mag \\
m444mag & F444W NIRCam magnitude\tnote{\textit{c}} \\
e\_m444mag & Error in m444mag\tnote{\textit{d}}\\
\hline
\end{tabular}
\begin{tablenotes}
\item[\textit{a}] Sources of the right ascension and declination are NIRCam
images from D25 reduced by L25, Hubble Space Telescope images from 
\citet{rob24} after alignment to NIRCam, or the APT catalog for JWST program 
5409.
\item[\textit{b}] Uncertainties are 0.5~subclass unless indicated otherwise.
\item[\textit{c}] Entries of ``sat" and ``out" indicate that the source was 
saturated or outside of the NIRCam image.
\item[\textit{d}] Synthetic photometric measurements in F444W from NIRSpec
are indicated by their assigned error, which is 0.13 mag.
\end{tablenotes}
\end{threeparttable}
\end{table}

\begin{figure}
\includegraphics[width=\columnwidth]{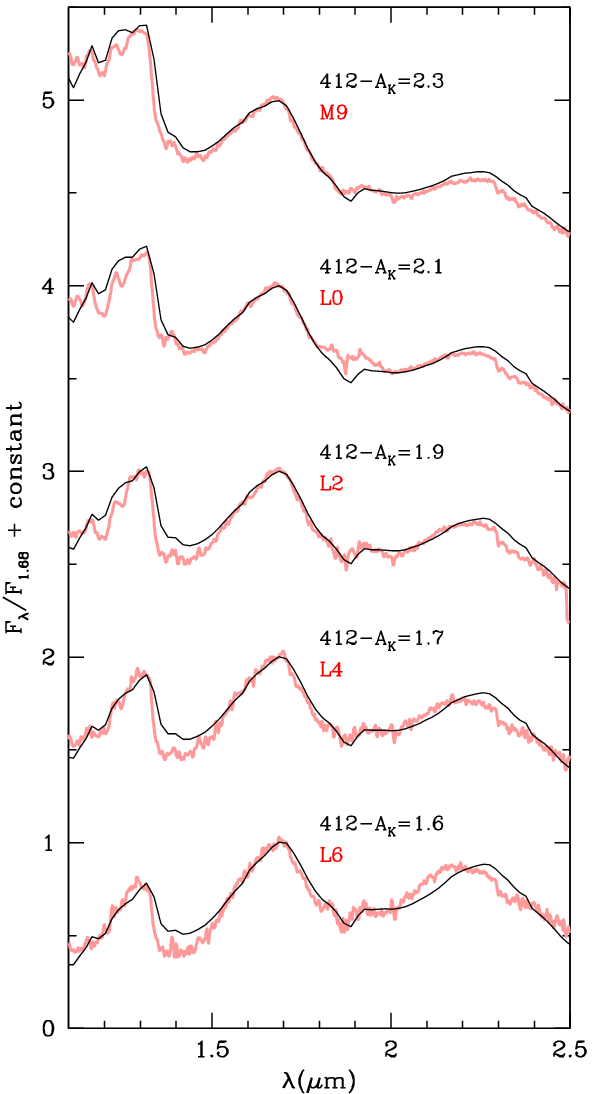}
\caption{NIRSpec spectrum of source 412 in NGC 2024 (black) dereddened to match
the 1--2.5~\micron\ slopes of young standards from M9--L6 \citep[red,][]{luh17}.
All of the standards are roughly consistent with source 412, illustrating
the degeneracy between spectral type and reddening for young L dwarfs.}
\label{fig:spec2}
\end{figure}

\section{Spectral Classifications}
\label{sec:class}

I have assessed whether each NIRSpec target is a likely member of NGC~2024
based on absorption features in the NIRSpec data that are sensitive to age
and other evidence of youth in the form of resolved circumstellar
structures in the NIRCam images (e.g., silhouette disks), excess 
emission from disks in the NIRSpec data or previous mid-IR photometry
\citep{kuh13}, X-ray emission \citep{bro13}, and emission lines
in the NIRSpec data. The spectral indicators 
of youth for late-type objects include triangular $H$-band continua and weak
CO absorption at 4.4--5.2~\micron\ \citep{luc01,luh23}. 
These indicators were identified through visual inspection of
both the observed spectra and the versions that were dereddened to match 
spectral standards. Sources with higher extinctions
tended to have lower signal-to-ratios (S/Ns) in the $H$ band, but
nearly all sources with late-type spectral features had adequate S/Ns 
for measuring the shape of the $H$-band continuum. Through this analysis,
I have classified 45 NIRSpec targets as likely members of NGC 2024 
($m_{444}\sim8$--18).
Among the remaining objects that have undetermined membership, several have
M spectral types (mostly indicative of low-mass stars) and positions in the 
Hertzsprung-Russell (H-R) diagram that are consistent with membership 
(Section~\ref{sec:hr}). The 28 sources that lack membership evidence
and late-type spectral features are probably background stars and galaxies.
Ten objects have undetermined membership because their NIRSpec data 
have insufficient wavelength coverage.

\begin{figure}
\includegraphics[width=\columnwidth]{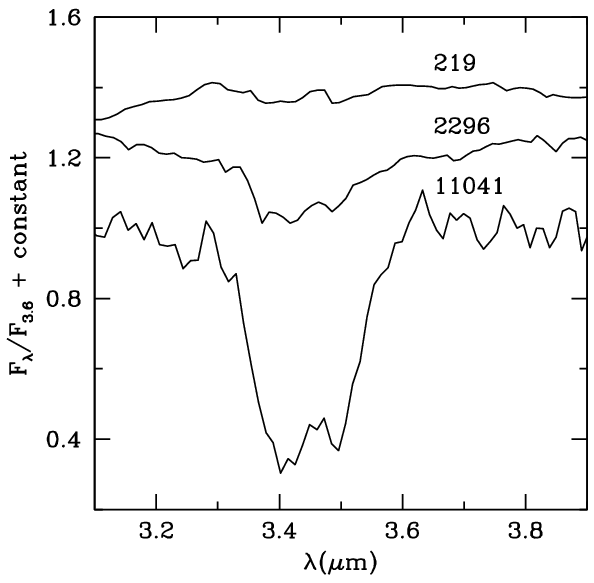}
\caption{NIRSpec spectra for source 219 in NGC 2024 and two brown dwarfs
in IC 348 (LRL 2296 and LRL 11041) that exhibit the 3.4~\micron\ band (L25).
Source 219 has a tentative detection of weak absorption in that band.
Its spectrum has S/N$\sim$300 according to the data reduction pipeline.}
\label{fig:spec3}
\end{figure}

For objects that have M/L-type spectral features, I have measured 
spectral types and extinctions through comparison of the NIRSpec data at
$<2.5$~\micron\ to standard spectra for young stars and brown dwarfs 
\citep[M0--L7,][]{luh17}. 
Stars across a wide range of spectral types exhibit similar spectral
slopes at $>1.5$~\micron, so I also have estimated extinctions
for the targets that lack M/L spectral types (both members and likely
background objects) as long as they have sufficient wavelength 
coverage and do not show IR excesses within NIRSpec's wavelength range. 
During the spectral classification
and other analysis in this study involving reddening, I have adopted
the extinction law from \citet{sch16av} for $x=-0.015$, which best
reproduces the reddening relations implied by NIRCam data in NGC 2024,
IC 348, and the ONC \citep[][L25]{luh24o2} when considering recent
extinction laws from 1--5~\micron\ \citep[e.g.,][]{gor23,wan24}.
Seventeen targets classified as members have spectral types indicative of
substellar masses for the age of NGC 2024 \citep[$>$M6,][]{bar15}.
The 87 NIRSpec targets are presented in Table~\ref{tab:spec},
which includes the spectral and membership classifications and photometry 
in five NIRCam filters on the Vega system, as measured in L25. 
The source numbers in Table~\ref{tab:spec} are from the 
catalog utilized by program 5409 in the Astronomer's Proposal Tool (APT).

In Figure~\ref{fig:spec1}, I present examples of spectra of sources that
are classified as members of NGC 2024, consisting of four M-type objects
and four stars that lack late-type features.  Sources 492 and 496 exhibit 
emission in the fundamental band of CO at 4.4--5.2~\micron. Source 93 is
the reddest NIRSpec target and is probably a protostar.

Among the adopted young standards, the H$_2$O bands become 
stronger from M0--M9 and deepen only slightly (1.3--1.6~\micron) or 
remain unchanged (1.8--2.1~\micron) from M9 to late L \citep{luh17}.
At low spectral resolution, the primary change across the latter range 
of types consists of the spectral slope from 1--2.5~\micron\ becoming
significantly redder. As a result, an early L dwarf with high extinction
can appear similar to a late L dwarf with low extinction \citep{luh17}, 
which means that L dwarfs in molecular clouds are subject to a 
degeneracy between spectral type and extinction. That degeneracy 
is illustrated in Figure~\ref{fig:spec2}, which compares the spectrum of
source 412 to young standards from M9--L6 for their best-fitting 
extinctions. Thus, the coolest objects in the NIRSpec sample have large ranges 
of possible spectral types. I note that the L2--L6 standard 
spectra in Figure~\ref{fig:spec2} provide worse fits to the H$_2$O bands 
in the NGC 2024 object than M9 and L0, but that could be due to the older ages 
of the L standards \citep[$\sim$10--100~Myr,][]{cru09}. For instance, 
as with source 412 in Figure~\ref{fig:spec2}, the L standards have stronger
H$_2$O bands than the young L-type companion TWA~27B \citep[10~Myr,][]{luh23}. 

NIRSpec observations of brown dwarfs in IC 348 have detected an absorption
band at 3.4--3.5~\micron\ from an unidentified aliphatic hydrocarbon
\citep[][L25]{luh24ic}, which is the basis of a proposed new spectral
class of ``H" (L25). Source 219 in NGC 2024 exhibits a tentative detection
of weak absorption in that band, as shown Figure~\ref{fig:spec3}.
That detection appears to have the same double dip structure that
has been found in the 3.4~\micron\ features in IC 348.
Source 219 is located outside of the NIRCam images
and it lacks mid-IR photometry from other telescopes. In IC~348, the brown 
dwarfs showing the 3.4~\micron\ band have $M_{444}\gtrsim7.7$. Only two of 
the brown dwarfs in the NIRCam images of NGC 2024 extend down to that 
magnitude range, so the absence of additional detections of the 
3.4~\micron\ band is not surprising.
Three of the brown dwarf candidates in NGC 2024 identified by L25 
lack spectroscopy and are faint enough that they could have strong
3.4~\micron\ bands (Section~\ref{sec:phot}).

Some NIRSpec targets have sufficiently high extinctions that the photospheric 
bands at shorter wavelengths are poorly measured, contributing to large 
uncertainties in the spectral types.
For objects with large uncertainties in spectral types, the extinctions 
in Table~\ref{tab:spec} apply to a spectral type that is one subclass later 
than the lower limit (e.g., L0 for a classification of M9--L).

\begin{figure}
\includegraphics[width=\columnwidth]{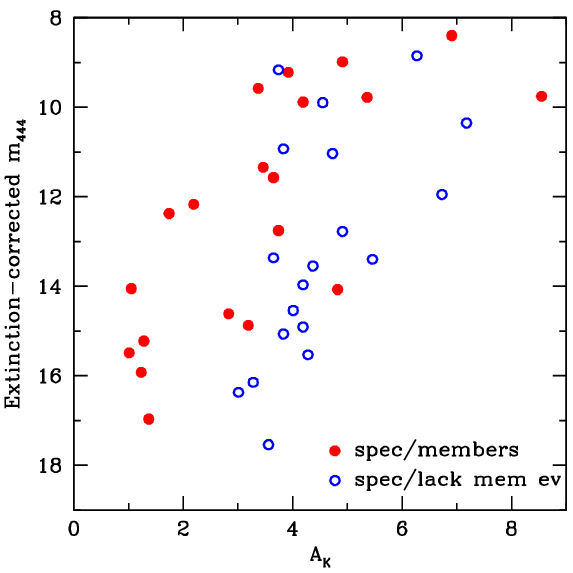}
\caption{Extinction-corrected $m_{444}$ versus $A_K$ for NIRSpec targets 
in NGC 2024, which are plotted with symbols that indicate whether they have
evidence of membership in the cluster.  For sources that are saturated in the 
NIRCam images, $m_{444}$ has been estimated from the NIRSpec data.}
\label{fig:ak} 
\end{figure}

To illustrate the range of extinctions for the NIRSpec targets, I have
plotted their extinction estimates and extinction-corrected values of $m_{444}$ 
in Figure~\ref{fig:ak} when both measurements are available. 
The NIRSpec targets with extinction estimates have minimum and median 
extinctions of $A_K\sim1$ and 3.7, respectively. Some of the reddest NIRSpec 
targets, such as source 93 from Figure~\ref{fig:spec1}, are absent 
from Figure~\ref{fig:ak} because the presence of IR excess emission
prevents reliable extinction estimates.

To include in Figure~\ref{fig:ak} stars that are saturated in the F444W images,
I have estimated photometry in that band from the NIRSpec spectra after scaling 
them to match the NIRCam photometry at shorter wavelengths. That synthetic 
photometry is subject to large errors because of wavelength-dependent slit 
losses in the NIRSpec data.  I estimate a typical error of $\sim$~0.13 mag
in the synthetic values of $m_{444}$ based on the standard deviation
of the differences between NIRCam and NIRSpec photometry for NIRSpec targets 
that are not saturated in F444W. The synthetic photometry in $m_{444}$ is 
included in Table~\ref{tab:spec}.

\begin{figure}
\includegraphics[width=\columnwidth]{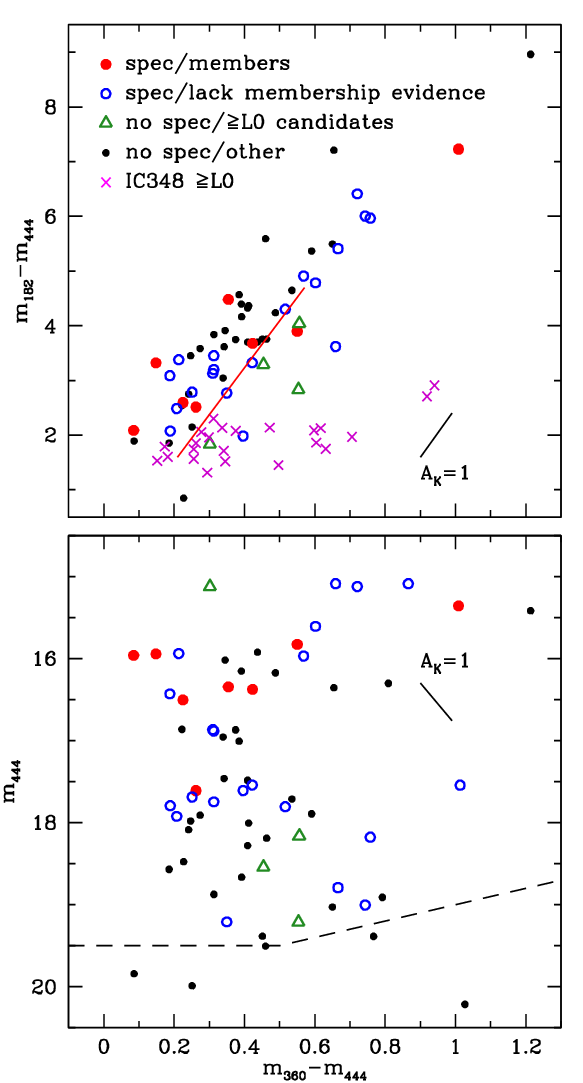}
\caption{Color-color and color-magnitude diagrams for sources in NIRCam
images of NGC 2024, which are plotted with symbols that indicate whether
they have NIRSpec spectra from this work, and if so, whether they have evidence
of membership in the cluster. Among sources that lack spectroscopy,
those that appear below the red reddening vector in the top diagram
are marked as candidates for brown dwarfs with types of $\geq$L0.
The known $\geq$L0 members of IC 348 are included in the top diagram
for reference (L25). The 50\% completeness limits in F360M and F444W are 
indicated by the dashed line.}
\label{fig:cmd}
\end{figure}

\section{Testing Photometric Identification of Brown Dwarf Candidates}
\label{sec:phot}

L25 used color-color and color-magnitude diagrams constructed from the
NIRCam photometry in NGC 2024 to identify brown dwarf candidates.
I revisit that analysis in light of the new spectral classifications from
NIRSpec.

The short/long wavelength (SW/LW) channels of NIRCam 
cover two $2\farcm2\times2\farcm2$ fields that are separated by $43\arcsec$. 
Each of those fields is imaged by four detectors separated by $5\arcsec$
for SW filters and one detector for LW filters. The NIRCam images in NGC 2024 
were taken at a single pointing using subpixel dithers.
Twenty of the NIRSpec targets are located outside of the NIRCam images,
and four additional targets appear within the gaps between SW detectors and
are saturated in the LW images. 
The remaining 63 NIRSpec targets have unsaturated detections in at least 
one NIRCam filter, many of which are saturated in the LW images.
Thus, only a minority of the NIRSpec targets appear in my analysis
of the NIRCam photometry (e.g., 29 have unsaturated data in both
F360M and F444W).

As discussed in L25, given the NIRCam bands available for NGC 2024,
diagrams of $m_{182}-m_{444}$ versus $m_{360}-m_{444}$ and
$m_{444}$ versus $m_{360}-m_{444}$ are the best options for identifying
brown dwarf candidates. In Figure~\ref{fig:cmd}, I have plotted those
diagrams with the photometry measured by L25, using different symbols
for the following categories of objects: 1) NIRSpec targets that I have
classified as members, 2) NIRSpec targets that have undetermined membership,
3) four NIRCam sources that lack spectroscopy and that were identified by L25 
as candidates for L-type brown dwarfs, and 4) the remaining sources that
lack spectra, excluding those that are extended based on a metric described
by \citet{luh24ic} and L25. 

Because the background emission varies significantly across the NIRCam
images, particularly in the LW filters, the sensitivity varies with
position as well. I have estimated the completeness limits of the 
F360M and F444W images by inserting artificial stars in the images
for a range of magnitudes and positions. I find that $\sim$50\% of the
artificial stars are recovered at $m_{360}\sim20$ and $m_{444}\sim19.5$.
Since some areas of background are very bright, a high completeness fraction
corresponds to much brighter magnitudes. I have marked the 50\% completeness
limits in the color-magnitude diagram in Figure~\ref{fig:cmd}. 

As shown in Figure~\ref{fig:cmd}, none of the sources classified as
members by NIRSpec approach the completeness limit. Instead, all of the
faintest NIRSpec targets ($m_{444}\gtrsim18$) lack late-type features and are 
likely to be background objects. Four of the eight brown dwarf candidates from 
L25 were observed by NIRSpec; one is confirmed as a young brown dwarf
(source 66 in this study), two are probably background stars, and
one (source 84) has insufficient wavelength coverage for classification 
(2.4--5.3~\micron), although it has photospheric H$_2$O absorption at 
2.5~\micron\ that is suggestive of an M/L spectral type.
Three of the four L25 candidates lacking spectra are fainter than
all of the confirmed brown dwarfs in the NIRSpec sample.
Spectroscopy of those candidates (and source 84) would be worthwhile
to determine if they are young brown dwarfs.
The candidates from L25 were identified based on their positions below
the red reddening vector in the color-color diagram in Figure~\ref{fig:cmd},
which captures most of the $\geq$L0 brown dwarfs in IC 348 (L25).
It is not possible to discriminate between brown dwarfs at earlier types
(M7--M9) and contaminants using the available bands of NIRCam photometry.

\section{Hertzsprung-Russell Diagram}
\label{sec:hr}

\begin{figure}
\includegraphics[width=\columnwidth]{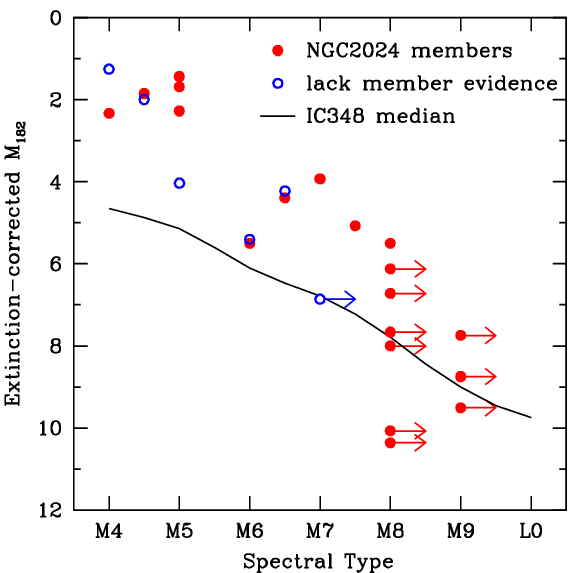}
\caption{Extinction-corrected $M_{182}$ versus spectral type for NIRSpec 
targets in NGC 2024.  The sources that lack evidence of membership have 
positions in this diagram that are roughly consistent with membership. 
For comparison, the median sequence for members of IC 348 is indicated 
\citep[$\sim5$~Myr,][]{luh16}.}
\label{fig:hr} 
\end{figure}

In Figure~\ref{fig:hr}, I have plotted an H-R diagram that consists
of extinction-corrected $M_{182}$ versus spectral type for
NIRSpec targets that have M/L spectral classifications, both cluster
members and sources with undetermined membership.
I have selected the F182M band because it offers unsaturated photometry 
for more of these objects than the LW filters.
As discussed in Section~\ref{sec:class}, the spectral types of some of the 
coolest sources have large uncertainties, spanning from late-M through L.
Those objects are plotted with arrows in Figure~\ref{fig:hr}.
The positions of the M-type objects that lack membership evidence are
roughly consistent with the sequence formed by the confirmed members.
For reference, I have marked the median sequence for IC 348, for which
I have adopted an age of 5~Myr \citep{luh24ic}.
The combination of that age for IC 348, the offset between the sequences
for NGC 2024 and IC 348 at M4--M6, and the fading rate 
predicted for young low-mass stars by \citet{bar15} suggests an age of 
$\sim0.5$~Myr for members of NGC 2024 within the NIRCam field, which 
is roughly consistent with previous age estimates \citep{lev06,get14b}.

\section{Initial Mass Function}
\label{sec:imf}

\begin{figure}
\includegraphics[width=\columnwidth]{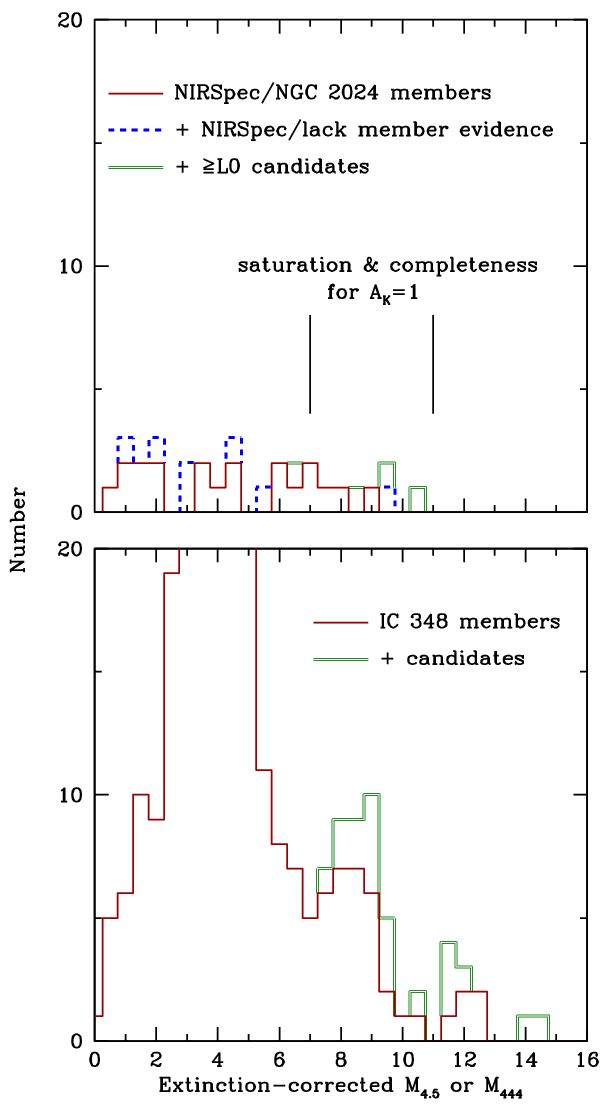}
\caption{
Histograms of extinction-corrected $M_{4.5}$ or $M_{444}$ for confirmed
and candidate members of NGC 2024 (top) and IC 348 (bottom, L25). 
The NGC 2024 sample is restricted to NIRSpec targets within the NIRCam 
field that have spectral classifications and the $\geq$L0 candidates in
those images, so it is incomplete for members that are unobserved by NIRSpec 
and are saturated in NIRCam.  The saturation and 50\% completeness limits for
the F444W data are marked for the minimum extinction of $A_K\sim1$ for members
observed by NIRSpec (Fig.~\ref{fig:ak}). Those limits shift to brighter 
magnitudes for higher extinctions. The NGC 2024 sample may be 
biased against members at lower masses because they are detected for a 
smaller range of extinctions (Fig.~\ref{fig:ak}). The sample for IC 348 is 
extinction-limited, so it should be unbiased in mass (L25).}
\label{fig:histo}
\end{figure}

To characterize the substellar IMF in NGC 2024, I use the F444W magnitude
as a proxy for luminosity, and in turn stellar mass, since it is less 
sensitive to extinction uncertainties than other available bands.
I have utilized a sample that consists of 
1) NIRSpec targets within the NIRCam field that have been classified as 
members in this work and that have extinction estimates from NIRSpec;
2) NIRSpec targets within the NIRCam field that have undetermined membership 
and measurements of spectral types and extinctions, which are promising
candidates for members (Fig.~\ref{fig:hr});
3) the four NIRCam brown dwarf candidates from L25 that lack spectroscopy;
and 4) source 84, which is a brown dwarf candidate from L25 that has uncertain
membership and spectral classifications because of incomplete NIRSpec 
coverage (Section~\ref{sec:phot}).
Figure~\ref{fig:histo} presents a histogram of extinction-corrected $M_{444}$
for these objects. For stars that are saturated in F444W, I have used
synthetic photometry in F444W derived from the NIRSpec data, as done
for Figure~\ref{fig:ak}.
Those synthetic estimates have large uncertainties, but they are adequate for 
the purpose of constructing the histogram, and are not relevant to the analysis 
of IMF at the lowest masses. 
For the five objects that lack extinction estimates from NIRSpec,
(source 84 and the four photometric candidates from L25), I have estimated
extinctions by dereddening their positions in the color-color diagram
in Figure~\ref{fig:cmd} to the colors of brown dwarfs in IC 348, arriving
at $A_K\sim0.5$ for source 84 and $A_K\sim1$, 3, 2, and 1.5 for
sources 1, 4, 5, and 7 from Table~3 in L25, respectively.
In Figure~\ref{fig:histo}, I have marked the
values of $M_{444}$ that correspond to saturation and 50\% completeness
(Section~\ref{sec:phot}) for the minimum extinction of $A_K\sim1$ 
for members observed by NIRSpec (Fig.~\ref{fig:ak}). Those limits 
shift to brighter $M_{444}$ for higher extinctions.

The NIRCam field encompasses $\sim$100 likely cluster members identified
in X-rays \citep{bro13} that were not observed by NIRSpec.
All of those objects are saturated in F444W, although a majority
have measurements in a similar band from the Spitzer Space Telescope
\citep{kuh13}. However, only a minority of the stars have the near-IR 
photometry or spectroscopy needed for extinction estimates. As a result, 
those stars are absent from the histogram in Figure~\ref{fig:histo}, which
means that the histogram is highly incomplete above the saturation level 
for NIRCam.

\begin{figure}
\includegraphics[width=\columnwidth]{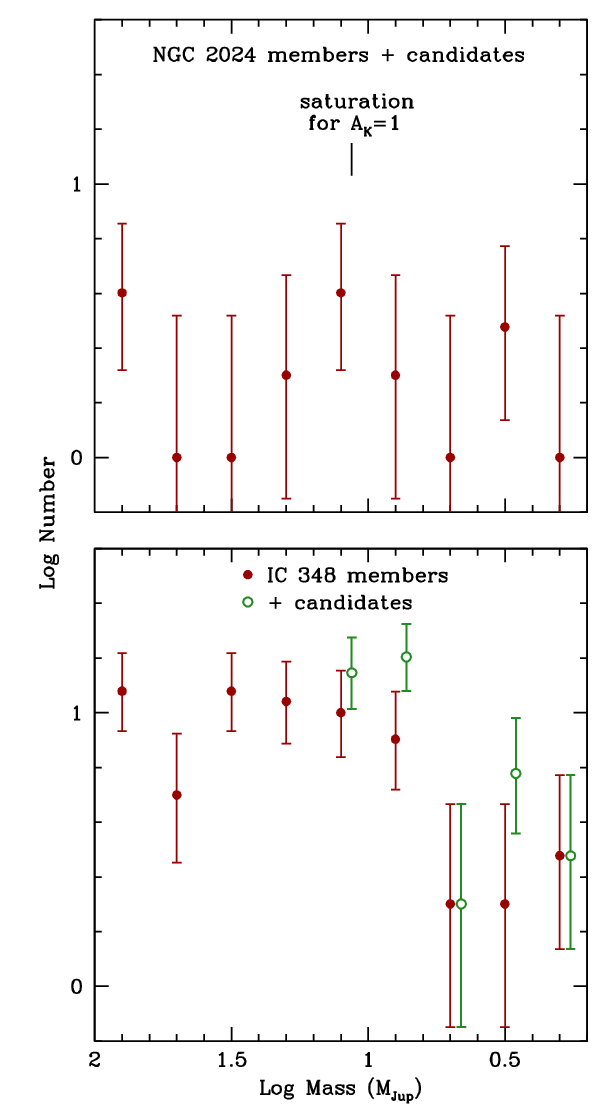}
\caption{Substellar IMFs for confirmed and candidate members of NGC 2024 (top) 
and IC 348 (bottom, L25).
The points for the IC 348 sample that include candidates are shifted slightly 
in mass for clarity.
}
\label{fig:imf}
\end{figure}

In general, more luminous members of a star-forming region are detectable
through higher levels extinctions. As a result, a sample of stars
from a magnitude-limited membership survey is likely to be biased against 
stars at lower masses, which are detected only at lower extinctions. 
That bias is evident in the diagram of extinction-corrected $m_{444}$ 
versus $A_K$ for the NIRSpec targets in NGC 2024 (Fig.~\ref{fig:ak}).
One can attempt to derive a mass function that is representative of the 
stellar population and is unbiased in mass by considering members within 
an extinction limit that provides an optimal combination of a large number 
of members (higher extinction limit) and a completeness limit that reaches 
low masses (lower extinction limit). However, the sample spanning all levels
of extinction (Figure~\ref{fig:histo}) is already quite small, and applying 
an extinction limit would reduce the sample size further. It is acceptable 
for my purposes to forgo an extinction threshold since my sample is incomplete 
above the saturation level of NIRCam and I am focusing on an IMF measurement
at the lowest masses that are probed by NIRCam.
For reference, I have included in Figure~\ref{fig:histo} an extinction-limited
sample of spectroscopically observed members of IC 348 and photometric
candidates that are located within a field imaged by NIRCam (L25).
That histogram employs 4.5~\micron\ photometry from Spitzer for sources
that are saturated in the F444W images from NIRCam.

As shown in Figure~\ref{fig:histo}, the histogram of NGC 2024 members and
photometric candidates extends down to the 50\% completeness limit,
and thus shows no detection of a minimum mass in the IMF.
Even if the photometric candidates were absent, the small size of the
sample would preclude a robust measurement of a minimum mass.

I have estimated masses for the objects in my
sample using their extinction-corrected $M_{444}$ and the values
predicted by evolutionary models for an age of 0.5~Myr (Section~\ref{sec:hr}).
I have utilized the models from \citet{bar15} and \citet{cha23} above
and below 10~$M_{\rm Jup}$, respectively. The youngest age included
in the latter models is 1~Myr, so I have extrapolated those model
predictions to 0.5~Myr by applying the $M_{444}$ offset between 0.5 and 1~Myr 
at 10~$M_{\rm Jup}$ from \citet{bar15}. The accuracy of evolutionary models 
is untested at very young ages and low masses, so these mass estimates 
may have substantial systematic errors. Among the NIRSpec targets 
confirmed as young and cool, the one with the faintest extinction-corrected 
$M_{444}$ (source 9) has a mass estimate of $\sim$4~$M_{\rm Jup}$. 
The faintest photometric candidate (source 7 from Table~3 in L25) has 
an estimate of $\sim$2~$M_{\rm Jup}$.

Using the mass estimates, I have constructed an IMF for the confirmed
and candidate members of NGC 2024 that were compiled earlier in this section.
The substellar portion of that IMF is shown in Figure~\ref{fig:imf}.
The masses in that IMF are presented in Table~\ref{tab:imf}.
I have marked the mass that corresponds to the NIRCam saturation limit in 
F444W for $A_K=1$. The F444W completeness limit for $A_K=1$ corresponds to
$\sim$2~$M_{\rm Jup}$. The mass function is plotted with logarithmic 
mass bins, which results in a Salpeter slope of 1.35. The statistical errors 
are from \citet{geh86}. In Figure~\ref{fig:imf}, I have included 
the IMF for the extinction-limited sample of members of IC~348 that was
shown in Figure~\ref{fig:histo} (L25). As in IC 348, spectroscopic 
surveys in other young clusters and associations have found substellar IMFs 
that are roughly flat down to 10~$M_{\rm Jup}$ \citep{zap17,dam23,luh25u}.

\begin{table}
\centering
\caption{Mass estimates for objects in the substellar IMF for NGC 2024
(Figure~\ref{fig:imf}). The format and content are described here.  
The full table is available in the online supplemental material.}
\label{tab:imf}
\begin{threeparttable}
\begin{tabular}{ll}
\hline
Column Label & Description\\
\hline
ID-5409 & Source number in catalog for JWST program 5409\\
ID-L25 & Source number from Table 3 in L25\\
RAdeg & Right ascension (ICRS)\tnote{\textit{a}}\\
DEdeg & Declination (ICRS)\tnote{\textit{a}}\\
mass & Mass estimated from $M_{444}$ \\
\hline
\end{tabular}
\begin{tablenotes}
\item[\textit{a}] Source of right ascension and declination is NIRCam images
from D25 reduced by L25.
\end{tablenotes}
\end{threeparttable}
\end{table}

Based on photometric candidates in the NIRCam data for NGC 2024, D25 reported 
the detection of a turnover in the IMF below 12~$M_{\rm Jup}$, and
found that the IMF was best fit by different power law slopes above and below
that mass. In Figure~\ref{fig:imf}, a turnover of that kind is not evident,
and the statistical constraints are too poor for identifying a change in
the shape of the IMF as a function of mass. In addition, as mentioned in
the discussion of the histogram of $M_{444}$, the IMF in NGC 2024
has incompleteness at higher substellar masses ($>10$~$M_{\rm Jup}$)
due to the saturation limit of the NIRCam data. The results from D25 utilized
28 candidates with mass estimates of 3--60~$M_{\rm Jup}$. In that mass range,
my IMF contains 14 objects, which suggests that the sample from D25 had 
significant contamination. It is not possible to make a direct comparison
of the two samples since D25 did not tabulate their candidates.
They identified brown dwarf candidates down to masses of 3~$M_{\rm Jup}$
and reported a sensitivity to objects below 1~$M_{\rm Jup}$, which they cited
as a possible detection of the minimum mass of the IMF.
However, as discussed in the context of the histogram of $M_{444}$, the 
faintest and least massive candidates in my sample (2--3~$M_{\rm Jup}$) 
coincide with the 50\% completeness limit for $A_K=1$, so I find no evidence
of a low-mass cutoff in the existing data. Even if the faintest photometric
candidates are rejected by future spectroscopy, the number of remaining 
confirmed brown dwarfs would be too small for a robust detection of a 
minimum mass.

\section{Conclusions}

I have analyzed low-resolution 1--5~\micron\ spectra of 87 sources toward
NGC 2024 that were obtained with JWST/NIRSpec through program 5409 
(PI: M. De Furio). Sixty-seven of the NIRSpec targets appear within
JWST/NIRCam images from D25. Based on those spectra and data from previous 
studies (e.g., X-ray emission), I have classified 45 of the NIRSpec targets 
as members of NGC 2024, 17 of which have spectral types that should
correspond to substellar masses (M6.5--L; 13 are within NIRCam). 
The coolest brown dwarfs have uncertain spectral types ($\sim$M8--L)
because of a degeneracy between spectral type and reddening.
One of the brown dwarfs exhibits a tentative detection of 
the 3.4~\micron\ band that has been observed in IC 348 brown dwarfs
and attributed to an unidentified aliphatic hydrocarbon \citep[][L25]{luh24ic}.
Some of the targets classified as members lack photospheric absorption
features for M/L types, which may be protostars or stars earlier than M,
and some lack spectral classifications because their spectra have
insufficient wavelength. Among the targets that lack 
membership evidence, some have M spectral types and are candidates for cluster 
members. The NIRSpec targets with extinction estimates have minimum and 
medium extinctions of $A_K\sim1$ and 3.7, respectively. Many of the reddest 
targets lack extinction estimates because of the presence of IR excess
emission (e.g., protostars).

I have used the NIRSpec classifications in conjunction with the photometry 
from NIRCam to constrain the substellar IMF in NGC 2024.
For the confirmed and candidate members observed by NIRSpec and four
photometric brown dwarf candidates from NIRCam that lack spectra, 
I have estimated masses from extinction-corrected $M_{444}$ and evolutionary 
models assuming a cluster age of 0.5 Myr. The faintest confirmed brown
dwarf has a mass estimate of $\sim$4~$M_{\rm Jup}$. The photometric candidates
extend down to the 50\% completeness limit in F444W ($\gtrsim2$~$M_{\rm Jup}$ 
for $A_K\gtrsim1$; Fig.~\ref{fig:histo}), so the minimum mass of the IMF in 
NGC 2024 has not been detected. Spectroscopy of those candidates would be
useful since their confirmation would provide a new limit on the IMF's 
minimum mass in NGC 2024. However, a robust measurement of 
a low-mass cutoff is not possible in the NIRCam field because of the small
number of brown dwarfs that it encompasses (Fig.~\ref{fig:imf}). 
Meanwhile, a reliable measurement of the shape of the substellar IMF in
this field is precluded by the small sample size as well as incompleteness at 
$\gtrsim$10~$M_{\rm Jup}$ due to the saturation limit of the NIRCam images
and a bias against brown dwarfs at lower masses due to the high extinctions
in the cluster. As a result, there is no evidence for the IMF turnover
below 12~$M_{\rm Jup}$ that was reported by D25.

\section*{Acknowledgements}

This work is based on observations made with the NASA/ESA/CSA 
James Webb Space Telescope. The data are associated with program 5409
and were obtained from MAST at the Space Telescope Science Institute, 
which is operated by the Association of Universities for Research 
in Astronomy, Inc., under NASA contract NAS 5-03127. 
The Center for Exoplanets and Habitable Worlds is supported by the
Pennsylvania State University, the Eberly College of Science, and the
Pennsylvania Space Grant Consortium.

\section*{Data Availability}

The data in Tables~\ref{tab:spec} and \ref{tab:imf} and the JWST spectra for 
those sources are available in the online supplemental material. 

\bibliographystyle{mnras}
\bibliography{ref} 

\bsp	
\label{lastpage}
\end{document}